# Symbian 'vulnerability' and Mobile Threats

Wajeb Gharibi

Head of Computer Engineering &Networks Department, Computer Science & Information Systems College,
Jazan University,
Jazan 82822-6694, Kingdom of Saudi Arabia
gharibi@jazanu.edu.sa

**Abstract**

Modern technologies are becoming ever more integrated with each other. Mobile phones are becoming increasing intelligent, and handsets are growing ever more like computers in functionality. We are entering a new era - the age of smart houses, global advanced networks which encompass a wide range of devices, all of them exchanging data with each other. Such trends clearly open new horizons to malicious users, and the potential threats are self evident.

In this paper, we study and discuss one of the most famous mobile operating systems 'Symbian'; its vulnerabilities and recommended protection technologies.

1. **Introduction**

Nowadays, there is a huge variety of cyber threats that can be quite dangerous not only for big companies but also for an ordinary user, who can be a potential victim for cybercriminals when using unsafe system for entering confidential data, such as login, password, credit card numbers, etc.

Modern technologies are becoming ever more integrated with each other. Mobile phones are becoming increasing intelligent, and handsets are growing ever more like computers in functionality. And smart devices, such as PDAs, on-board car computers, and new generation household appliances are now equipped with communications functions. We are entering a new era - the age of smart houses, global networks which encompass a wide range of devices, all of them exchanging data with each other via - as cyberpunk authors say - air saturated with bits and bytes. Such trends clearly open new horizons to malicious users, and the potential threats are self evident.

Our paper is organized as follows: Section 2 demonstrates the mobile operating system 'Symbian' vulnerabilities. Section3 proposes Symbians' Trojan Types. Section 4 recommends some possible protection techniques. Conclusions have been made in Section 5.

**2. Symbian Vulnerabilities**

The term 'vulnerability' is often mentioned in connection with computer security, in many different contexts. It is associated with some violation of a security policy. This may be due to weak security rules, or it may be that there is a problem within the software itself. In theory, all types of computer/mobile systems have vulnerabilities [1-5].

Symbian OS was originally developed by Symbian Ltd.[4]. It designed for smartphones and currently maintained by Nokia. The Symbian platform is the successor to Symbian OS and Nokia Series 60; unlike Symbian OS, which needed an additional user interface system, Symbian includes a user interface component based on S60 5th Edition. The latest version, Symbian^3, was officially released in Q4 2010, first used in the Nokia N8.

Devices based on Symbian accounted for 29.2% of worldwidesmartphone market share in 2011 Q1.[5] Some estimates indicate that the cumulative number of mobile devices shipped with the Symbian OS up to the end of Q2 2010 is 385 million [6].

On February 11, 2011, Nokia announced a partnership with Microsoft which would see it adoptWindows Phone 7 for smartphones, reducing the number of devices running Symbian over the coming two years.[12]

Symbian OS was subject to a variety of viruses, the best known of which is Cabir. Usually these send themselves from phone to phone by Bluetooth. So far, none have taken advantage of any flaws in Symbian OS – instead, they have all asked the user whether they would like to install the software, with somewhat prominent warnings that it can't be trusted.

This short history started in June 2004, when a group of professional virus writers known as 29A created the first virus for smartphones. The virus called itself 'Caribe'. It was written for the Symbian operating system, and spread via Bluetooth. Kaspersky Lab classified the virus as Worm.SymbOS.Cabir.

Although a lot of media hype surrounded Worm.SymbOS.Cabir, it was actually a proof of concept virus, designed purely to demonstrate that malicious code could be created for Symbian. Authors of proof of concept code assert that they are motivated by curiosity and the desire to improve the security of whichever system their creation targets; they are therefore usually not interested either in spreading their code, or in using it maliciously. The first sample of Cabir was sent to antivirus companies at the request of its author. The source code of the worm was, however, published on the Internet, and this led to a large number of modifications being created. And because of this Cabir started too slowly but steadily infect telephones around the world.

A month after Cabir appeared, antivirus companies were startled by another technological innovation: Virus.WinCE.Duts. It occupies a double place of honour in virus collections - the first known virus for the Windows CE (Windows Mobile) platform, and also the first file infector for smartphones. Duts infects executable files in the device's root directory, but before doing this, requests permission from the user.

A month after Duts was born, Backdoor.WinCE.Brador made its appearance. As its name shows, this program was the first backdoor for mobile platforms. The malicious program opens a port on the victim device, opening the PDA or smartphone to access by a remote malicious user. Brador waits for the remote user to establish a connection with the compromised device.

With Brador, the activity of some of the most experienced in the field of mobile security - the authors of proof of concept viruses, who use radically new techniques in their viruses - comes almost to a standstill. Trojan.SymbOS.Mosquit, which appeared shortly after Brador, was presented as Mosquitos, a legitimate game for Symbian, but the code of the game had been altered. The modified version of the game sends SMS messages to telephone numbers coded into the body of the program. Consequently, it is classified as a Trojan as it sends messages without the knowledge or consent of the user - clear Trojan behaviour.

In November 2004, after a three month break, a new Symbian Trojan was placed on some internet forums dedicated to mobiles. Trojan.SymbOS.Skuller, which appeared to be a program offering new wallpaper and icons for Symbian was an SIS file - installer for Symbian platform. Launching and installing this program on the system led to the standard application icons (AIF files) being replaced by a single icon, a skull and crossbones. At the same time, the program would overwrite the original applications which would cease to function.

Trojan.SymbOS.Skuller demonstrated two unpleasant things about Symbian architecture to the world. Firstly, system applications can be overwritten. Secondly, Symbian lacks stability when presented with corrupted or non-standard system files - and there are no checks designed to compensate for this 'vulnerability'.

This 'vulnerability' was quickly exploited by those who write viruses to demonstrate their programming skills. Skuller was the first program in what is currently the biggest class of malicious programs for mobile phones. The program's functionality is extremely primitive, and created simply to exploit the peculiarity of Symbian mentioned above. If we compare this to PC viruses, in terms of damage caused and technical sophistication, viruses from this class are analogous to DOS file viruses which executed the command 'format c:\' .

The second Trojan of this class - Trojan.SymbOS.Locknut - appeared two months later. This program exploits the trust shown by the Symbian developers (the fact that Symbian does not check file integrity) in a more focused way. Once launched, the virus creates a folder called 'gavno' (an unfortunate name from a Russian speaker's point of view) in /system/apps. The folder contains files called 'gavno.app', 'gavno.rsc' and 'gavno_caption.rsc'. These files simply contain text, rather than the structure and code which would normally be found in these file formats. The .app extension makes the operating system believe that the file is executable. The system will freeze when trying to launch the application after reboot, making it impossible to turn on the smartphone.

### 3. Symbians' Trojan Types

Trojans exploiting the Symbian 'vulnerability' differ from each other only in the approach which is used to exploit the 'vulnerability'.

a) Trojan.SymbOS.Dampig overwrites system applications with corrupted ones

b) Trojan.SymbOS.Drever prevents some antivirus applications from starting automatically

c) Trojan.SymbOS.Fontal replaces system font files with others. Although the replacement files are valid, they do not correspond to the relevant language version of the font files of the operating system, and the result is that the telephone cannot be restarted

d) Trojan.SymbOS.Hoblle replaces the system application File Explorer with a damaged one

e) Trojan.SymbOS.Appdiasbaler and Trojan.SymbOS.Doombot are functionally identical to Trojan.SymbOS.Dampig (the second of these installs Worm.SymbOS.Comwar)

f) Trojan.SymbOS.Blankfont is practically identical to Trojan.SymbOS.Fontal

The stream of uniform Trojans was broken only by Worm.SymbOS.Lascon in January 2005. This worm is a distant relative of Worm.SymbOS.Cabir. It differs from its predecessor in that it can infect SIS files. And in March 2005 Worm.SymbOS.Comwar brought new functionality to the mobile malware arena - this was the first malicious program with the ability to propagate via MMS.

**4. Possible Protection Techniques**

Mobile has security vulnerabilities like computer and network. There is no particular locking system or guarding system that is able to ensure 100 percent security. Conversely, there are various types of security locks or guards that are suitable for different situations. We can make use of the combination of available and up to date technologies to fight the serious attacks. Yet there is no guaranty that this option will provide 100 percent security, nevertheless, this methodology certainly maximizes the mobile security and it is often possible to stop a threat. Few techniques are documented here which are also suggested by Wi-Fi Planet, 2007; TechRepublic, 2008; and TechGuru, 2010.

- Enable SIM, device and access lock from mobile settings. Enable the periodic lockdown feature. Enable the memory access code.
- Think deeply before accessing any internet site and installing any application.
- Spend little bit more time to check the application through Google or any search engine before downloading or installing unknown files.
- Disable WLAN and Bluetooth when you are out door and when you are not using it.
- Find a phone with the service option to remotely kill it when it is irretrievably lost.
- Never let others access your phone. Be careful while accepting calls or messages from unknown numbers.
- Enable WPA2 encryption for WLAN connection and pass code request feature for Bluetooth connection.
- If you noticed that your phone has connected to GPRS, UMTS, and HSDPA, disable those instantly.
- Keep regular backup.
- Install antivirus software.
- Do not simply save sensitive information on the phone unless absolutely essential.

**5. Trends and forecasts**

It is difficult to forecast the evolution of mobile viruses with any accuracy. This area is constantly in a state of instability. The number of factors which could potentially provoke serious information security threats is increasing more quickly than the environment - both technological and social - is adapting and evolving to meet these potential threats.

The following factors will lead to an increase in the number of malicious programs and to an increase in threats for smartphones overall:

- The percentage of smartphones in use is growing. The more popular the technology, the more profitable an attack will be.
- Given the above, the number of people who will have a vested interested in conducting an attack, and the ability to do so, will also increase.
- Smartphones are becoming more and more powerful and multifunctional, and beginning to squeeze PDAs out of the market. This will offer both viruses and virus writers more functionalities to exploit.
- An increase in device functionality naturally leads to an increase in the amount of information which is potentially interesting to a remote malicious user that isstored on the device. In contrast to standard mobile phones, which usually have little more than an address book stored on them, a smartphone memory can contain any files which would normally be stored on a computer hard disk. Programs which give access to password protected online services such as ICQ can also be used on smartphones, which places confidential data at risk.

However, these negative factors are currently balanced out by factors which hinder the appearance of the threats mentioned above: the percentage of smartphones remains low, and no single operating system is currently showing dominance on the mobile device market. This currently acts as a brake on any potential global epidemic - in order to infect the majority of smartphones (and thus cause an epidemic) a virus would have to be multiplatform. Even then the majority of mobile network users would be secure as they would be using devices with standard (not smartphone) functionality.

Mobile devices will be under serious threat when the negative factors start to outweigh the positive. And this seems to be inevitable. According to data from the analytical group SmartMarketing, the market share of Symbian on the Russian PDA and smartphone market has been steadily increasing over the last 2 to 3 years. By the middle of 2005 it had a market share equal to that of Windows Mobile, giving rise to the possibility that the former may be squeezed out of the market.

Currently, there is no threat of a global epidemic caused by mobile malware. However, the threat may become real a couple of years down the line - this is approximately how long it will take for the number of smartphones, experienced virus writers and platform standardization to reach critical mass. Nevertheless, this does not reduce the potential threat - it's clear that the majority of virus writers are highly focussed on the mobile arena. This means that viruses for mobile devices will invariably continue to evolve, incorporating/ inventing new technologies and malicious payloads which will gradually become more and more widespread. The number of Trojans for Symbian which exploit the system's weak points will also continue to grow, although the majority of them are likely to be primitive (similar in functionality to Fontal and Appdisabler).

The overall movement of virus writers into the mobile arena is an equal stream of viruses analogous to those which are already known with the very rare inclusion of technological novelties and this trend seems likely to continue for the next 6 months at minimum. An additional stimulus for viruses writers will be the possibility of financial gain, and this will come when smartphones are widely used to conduct financial operations and for interaction with e-payment systems.

## 6. Conclusions

Smart mobile devices are still in their infancy, and consequently very vulnerable, both from a technical and a sociological point of view. On the one hand, their technical stability will improve only under arms race conditions, with a ceaseless stream of attacks and constant counter measures from the other side. This baptism of fire has only just begun for PDAs and smartphones, and consequently security for such devices is, as yet, almost totally undeveloped.